\documentclass[a4paper,10pt,twocolumn,superscriptaddress]{revtex4}
\usepackage{amsmath,amssymb}
\usepackage[final]{graphicx}

\begin{document}

\title{Transport anomaly at the ordering transition for adatoms on graphene}
\author{Sergey Kopylov}
\affiliation{Department of Physics, Lancaster University, Lancaster, LA1~4YB, UK}
\author{Vadim Cheianov}
\affiliation{Department of Physics, Lancaster University, Lancaster, LA1~4YB, UK}
\author{Boris L. Altshuler}
\affiliation{Physics Department, Columbia University, New York, N.Y. 10027, USA}
\author{Vladimir I. Fal'ko}
\affiliation{Department of Physics, Lancaster University, Lancaster, LA1~4YB, UK}

\begin{abstract}
We analyze a manifestation of the partial ordering transition of adatoms on graphene
in resistivity measurements. We find that Kekul\'{e} mosaic ordering of adatoms increases
sheet resistance of graphene, due to a gap opening in its spectrum, and that critical fluctuations
of the order parameter lead to a non-monotonic temperature dependence of resistivity, with a
cusp-like minimum at $T=T_c$.
\end{abstract}

\maketitle

Impurities in metals experience a long-range RKKY interaction due to polarization of the electron
Fermi sea (Friedel oscillations) \cite{FriedelOriginal}. For surface adsorbents such an interaction may result
in their structural ordering, repeating the pattern of the Friedel oscillations
of electron density \cite{FriedelOsc}. In particular, a dilute ensemble
of adatoms on graphene may undergo a partial ordering transition \cite{Kekule,AllTypes,Sublattice,Levitov}.

Unlike other materials, the RKKY interaction between
adatoms on graphene exists even at zero carrier density, with a characteristic long-range
$1/r^3$ dependence, and it exhibits the Friedel oscillations which are commensurate with the underlying honeycomb
lattice. For adatoms residing above the centers of the honeycombs,
the intervalley scattering of the electrons by adatoms leads to Friedel oscillations that resemble
a $\sqrt{3}\times\sqrt{3}$ charge-density wave superlattice.
Positions of each individual adatom
relative to such superlattice can be characterized by one of three
vectors, ${\bf u} = (\cos\frac{2\pi m}{3}, \sin\frac{2\pi m}{3})$, with $m=-1,0,1$.
The transition of an ensemble of adatoms into a ``Kekul\'{e} mosaic" ordered state \cite{Kekule},
characterized by the order parameter $\overline{\bf u}$, falls
in the symmetry class of 3-state Potts models \cite{Baxter}.

\begin{figure}[tbp]
\centering
\includegraphics[width=.96\columnwidth]{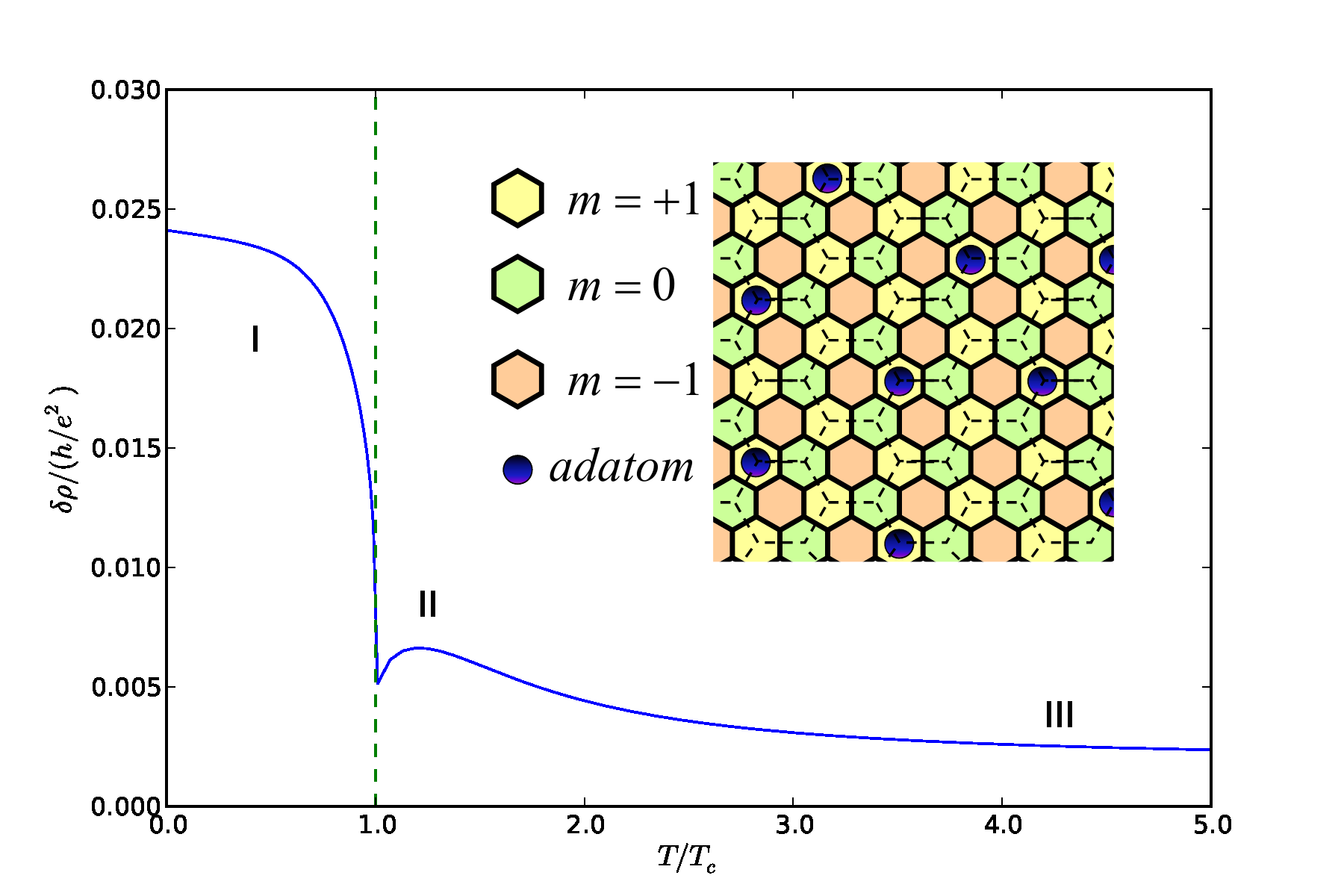}
\caption{The predicted anomaly in the temperature-dependent resistivity of graphene decorated
with adatoms in the vicinity of the Kekul\'{e} ordering transition. The inset illustrates the Kekul\'{e}
mosaic ordered state and the assignment of Potts ``spin" $m=-1,0,1$ to various hexagons in the
$\sqrt{3}\times\sqrt{3}$ superlattice.}
\label{fig:resist}
\end{figure}

In this paper we analyze how partial ``Kekul\'{e}" ordering of adatoms on graphene affects its resistivity, $\rho$,
in the regime of low coverage $n_i a^2\ll 1$ ($n_i$ is the concentration of adatoms, $a$ is the lattice constant).
The behaviour of the temperature dependent resistivity correction $\delta\rho(T)=\rho(T)-\rho(\infty)$
is sketched in Fig.~\ref{fig:resist}.
At $T\ll T_c$ (region I) the temperature dependence of the resistivity is dominated by
a non-vanishing order parameter $\overline{\bf u}$ causing the amplified intervalley mixing and opening
a gap $\Delta \propto \overline{\bf u}\sim(T_c-T)^{\beta}$ in the corners of the Brillouin zone.
As the temperature increases from $T=0$ to $T=T_c$ the resistivity correction monotonically decreases as $(T_c-T)^{2\beta}$.
At $T>T_c$, critical fluctuations of the order parameter, characterized by the correlation length
$\xi\propto |T-T_c|^{-\nu}$, preceding formation of the ordered phase lead to a non-monotonic feature in $\delta\rho(T)$. 
At high temperature $T\gtrsim T_c$ (region III), the constructive interference of electron waves, scattered by adatoms
within ordered clusters of size $\xi$, enhances resistivity. The effect, which becomes stronger upon approaching $T_c$,
is similar to the critical opalescence \cite{Opalescence} in materials
undergoing structural phase transition or resistivity anomaly in bulk metals with magnetic impurities
undergoing ferromagnetic transition \cite{FisherLanger}.
This enhancement saturates when $\xi$ becomes comparable
to the electron wavelength $\lambda_F$, $\lambda_F\approx\xi$. In the region II
of temperatures $T\to T_c+0$, where $\xi\gg\lambda_F$, scattering of electrons
is affected only by the gradient of the fluctuating order parameter $\overline{\bf u}$.
The resistivity is thus reduced and a cusp-shape minimum at $T=T_c$ should be expected.

In the following we assume that the electron concentration $n_e=4\pi/\lambda^2_F$ is not high, $n_e\ll n_i$,
but the electron Fermi wavelength is shorter than its mean free path, $\lambda_F\ll l$.
This assumption also implies that in the ordered phase $k_B T, \Delta\ll \varepsilon_F$.

The electrons are described by a four-component Dirac-like spinor
$\Psi = [\psi_{K, A}, \psi_{K, B}, \psi_{K', B}, \psi_{K', A}]$
with the components corresponding to different valleys ($K$, $K'$) and sublattices ($A$, $B$) \cite{McClure}.
In the absence of adatoms, quasiparticles are characterized by the linear spectrum $|\varepsilon_{\bf p}|=\hbar vp$
and plane wave states (for $\varepsilon_{\bf p}>0$)
\begin{equation}
|K{\bf p}\rangle = \frac{e^{i{\bf pr}}}{\sqrt{2S}}
\left( \begin{array}{c}
1 \\ e^{i\varphi_{\bf p}} \\ 0 \\ 0 \\
\end{array} \right), \quad
|K'{\bf p}\rangle = \frac{e^{i{\bf pr}}}{\sqrt{2S}}
\left( \begin{array}{c}
0 \\ 0 \\ 1 \\ -e^{i\varphi_{\bf p}} \\
\end{array} \right).
\nonumber
\end{equation}
Here $S$ is the area of the graphene sheet and ${\bf p}=(p\cos\varphi_{\bf p}, p\sin\varphi_{\bf p})$
is the electron wave vector.

The Hamiltonian describing graphene covered by adatoms has the form \cite{Kekule,WeakLocaliz}
\begin{eqnarray}
& \hat{H} = \hbar v({\bf p\Sigma}) + \hat{U}({\bf r}) + \hat{V}({\bf r}); \label{eq:hamilt} \\
& \hat{U} = \sum\limits_l \hat{I} w({\bf r-r}_l); \quad
\hat{V} = \hbar\lambda v a \sum\limits_l \Sigma_z ({\bf u}_l {\bf \Lambda}) \delta({\bf r}-{\bf r}_l).
\nonumber
\end{eqnarray}
We assume that the dimensionless impurity potential $\lambda$ is small ($\lambda\lesssim 1$)
and will treat the electron-adatom interaction perturbatively.
We use the set \cite{WeakLocaliz,FriedelOsc} of $4\times 4$ "sublattice" and "valley"
matrices $\Sigma_{x,y,z}$ and $\Lambda_{x,y,z}$
\begin{eqnarray}
& \Sigma_x = \Pi_z\otimes\sigma_x, \Sigma_y = \Pi_z\otimes\sigma_y, \Sigma_z = \Pi_0\otimes\sigma_z, \nonumber \\
& \Lambda_x = \Pi_x\otimes\sigma_z, \Lambda_y = \Pi_y\otimes\sigma_z, \Lambda_z = \Pi_z\otimes\sigma_0, \nonumber
\end{eqnarray}
where $\sigma_i$ and $\Pi_j$ are Pauli matrices in sublattice (AB) and valley ($KK'$) spaces.

The form of the electron-adatom interaction in Eq.~(\ref{eq:hamilt}) is prescribed by the highly symmetric position
of adatoms at the centers of hexagons. The $\hat{U}$-term does not violate the
AB sublattice symmetry and scatters electrons without changing their valley state. The $\hat{V}$-term
\cite{FriedelOsc,Kekule,AllTypes} is responsible for the intervalley scattering
of electrons. The sensitivity of the scattering phase of an electron to the position of the adatom in
the $\sqrt{3}\times\sqrt{3}$ superlattice manifests itself by the Potts parameter ${\bf u}$ in the intervalley term.

Each adatom creates Friedel oscillations of the electron density leading
to the RKKY-type interaction between adatoms. 
The contribution of the $\hat{U}$-term, Eq.~\ref{eq:hamilt}, to such an interaction is a
$1/r^3$ repulsion independent on the Potts spins.
At the same time the symmetry-breaking coupling leads to 3-state Potts model with a long-range interaction
$- {\bf u}_j \cdot {\bf u}_l /r^3_{jl}$ \cite{Kekule}.

To minimize the interaction energy adatoms have to take equivalent positions within the
$\sqrt{3}\times\sqrt{3}$ superlattice unit cells. A Monte Carlo simulation of the corresponding
$1/r^3$ random-bond Potts model \cite{Kekule} has revealed such an ordered phase below the transition temperature
$T_c\approx 0.6 \lambda^2 (n_i a^2)^{3/2} \hbar v/a$.

The effects of adatoms ordering on the electron transport are encoded in the correlation function,
\begin{equation}
\overline{u^\alpha({\bf r}) u^\beta({\bf r'})} - \overline{u^\alpha({\bf r})}\,\,
\overline{u^\beta({\bf r'})} = \delta_{\alpha\beta} g(|{\bf r-r'}|),
\nonumber
\end{equation}
for which the theory of critical phenomena predicts the scaling form \cite{Fisher}
\begin{equation}
g(r) = \frac{\kappa(r/\xi)}{(\sqrt{n_i}r)^\eta}, \quad
\kappa(y) = \kappa_1(y) + y^{\frac{1-\alpha}{\eta}}\kappa_2(y).
\label{eq:corr_fnc}
\end{equation}
Here, $\xi\sim n^{-1/2}_i |(T-T_c)/T_c|^{-\nu}$ is the correlation length, and $\eta$ and $\nu$ are critical exponents.
It has been shown \cite{Baxter} that for the
3-state Potts model on a square lattice $\eta=4/15$, however, the value of $\eta$
for random-bond Potts models with a long-range interaction is still unknown \cite{CriticalExp,Dotsenko}.
In the critical region $r\ll\xi$, the correlation function has essentially $r^{-\eta}$ behaviour,
with a correction (second term) related to the specific heat anomaly $C\sim |T-T_c|^{-\alpha}$ \cite{Baxter}.
At large distances, $r\gg\xi$, $g(r)$ decays exponentially, according to the Ornstein-Zernike theory \cite{LandauLifshic,Fisher}.
Overall, for $T>T_c$ the scaling functions $\kappa_{1,2}(y$) have the following asymptotics:
\begin{eqnarray}
& \kappa_1(y\ll 1) \approx c_1, \quad \kappa_2(y\ll1 ) \approx -c_2 \quad (c_1, c_2 \sim 1); \nonumber \\
& \kappa_1(y\gg 1) \propto \displaystyle \frac{e^{-y}}{y^{1/2-\eta}}, \quad \kappa_2(y\gg 1) \sim e^{-y},
\label{eq:correl_above}
\end{eqnarray}
whereas for $T<T_c$,
\begin{equation}
\begin{array}{l}
\kappa_1(y\ll 1) \approx c_1, \quad \kappa_2(y\ll 1) \approx c_2; \\
\kappa_1(y\gg 1) \propto \displaystyle \frac{e^{-y}}{y^{2-\eta}}, \quad \kappa_2(y\gg 1) \sim e^{-y}. \\
\end{array}
\nonumber
\end{equation}

At $T<T_c$, the order parameter also acquires a homogenious average $\overline{\bf u} \propto (T_c-T)^\beta$.
As long as $\bar{\bf u}$ exceeds fluctuations of ${\bf u}$
(i.e. far enough from $T_c$), the electron states, which wavelength
is larger than the distance between adatoms $n^{-1/2}_i$, are described by the effective mean-field Hamiltonian,
\begin{equation}
\overline{\hat{H}} = \hbar v{\bf p\Sigma} + n_i \hbar\lambda va \Sigma_z (\overline{\bf u} \Lambda).
\nonumber
\end{equation}
Accordingly the spectrum $\varepsilon^2_{\bf p} = (\hbar vp)^2+\Delta^2$ acquires a gap,
\begin{equation}
\Delta(T)\approx n_i \hbar\lambda va (1-T/T_c)^\beta,
\end{equation}
such that $\Delta(0)\gg T_c$.
The plane wave eigenstates of $\overline{\hat{H}}$ are mixed between the two valleys and take the form
(for $\varepsilon_{\bf p}>0$)
\begin{equation}
|\pm1,{\bf p}\rangle = \frac{e^{i{\bf pr}}}{\sqrt{4S}} \left( \begin{array}{c}
    \sqrt{\frac{\varepsilon_{\bf p}\pm\Delta}{\varepsilon_{\bf p}}} \\
    \sqrt{\frac{\varepsilon_{\bf p}\mp\Delta}{\varepsilon_{\bf p}}} e^{i\varphi_{\bf p}} \\
    \pm \sqrt{\frac{\varepsilon_{\bf p}\pm\Delta}{\varepsilon_{\bf p}}} e^{i\theta} \\
    \mp \sqrt{\frac{\varepsilon_{\bf p}\mp\Delta}{\varepsilon_{\bf p}}} e^{i(\varphi_{\bf p}+\theta)} \\
\end{array} \right),
\overline{u_x}+i \overline{u_y} = u e^{i\theta}.
\nonumber
\end{equation}

Intravalley and intervalley scattering determined by $\hat{U}({\bf r})$ and $\hat{V}({\bf r})$
in Eq.~(\ref{eq:hamilt}) respectively do not interfere with each other. Hence,
the total momentum relaxation rate is the sum of the two electron scattering rates,
\begin{equation}
\tau^{-1} = \tau^{-1}_0 + \tau^{-1}_i,
\label{eq:scatter-rate}
\end{equation}
where $\tau_0$ and $\tau_i$ stand for intravalley and intervalley momentum relaxation times.
For the temperature-dependent Drude resistivity of graphene sheet
(recall that $k_B T, \Delta \ll\varepsilon_F$) we thus have
\begin{equation}
\rho(T) = \frac{2}{e^2} \frac{1}{v^2_F\tau\nu},
\label{eq:rho}
\end{equation}
where $v_F = \hbar v^2 p_F/\varepsilon_F$ is the Fermi velocity,
$\nu=2\varepsilon_F/(\pi\hbar^2 v^2)$ is the density of states, and the Fermi energy and momentum
are related to the electron density as $p_F=\sqrt{\pi n_e}$ and $\varepsilon_F=\sqrt{\pi \hbar^2 v^2 n_e+\Delta^2}$.

The temperature dependence $\rho(T)$ at $T\lesssim T_c$ is dominated by the effect of the order parameter $\bar{\bf u}$ on
the chiral plane wave functions and thus on the scattering rates, in particular $\tau^{-1}_0$.
In the Born approximation
\begin{eqnarray}
& \displaystyle \frac{1}{\tau_0} = \frac{n_i p^2_F}{\hbar \varepsilon_F}
\int\limits^{2\pi}_0 \frac{d\varphi}{2\pi} \tilde{w}^2\left( 2p_F \sin\frac{\varphi}{2} \right)
(1-\cos\varphi) r_0(\varphi), \label{eq:tau_0} \\
& r_0(\varphi) = \displaystyle \cos^2\frac{\varphi}{2} + \frac{\Delta^2(T)}{(\hbar vp_F)^2}, \nonumber
\end{eqnarray}
where $\tilde{w}(k)=\int d{\bf r} \, e^{i{\bf kr}} w(r)$ and $\varphi$ is the scattering angle.
The form-factor $r_0(\varphi)$ arises from the overlap integral between plane wave states and reflects the
absence of the backscattering for $\Delta=0$. Thus, for $T\lesssim T_c$, we find
\begin{equation}
\frac{\delta\rho(T)}{\rho(\infty)} \approx \frac{4\Delta^2(T)}{\pi n_e\hbar^2 v^2}
\frac{\int^{2\pi}_0 d\varphi\, \tilde{w}^2\left(2p_F \sin\frac{\varphi}{2}\right) \sin^2\frac{\varphi}{2}}
{\int^{2\pi}_0 d\varphi\, \tilde{w}^2\left(2p_F \sin\frac{\varphi}{2}\right) \sin^2\varphi}.
\label{eq:rho_delta}
\end{equation}

The temperature dependence $\rho(T)$ at $T>T_c$ is determined by the effect of the ordering of adatoms on
the intervalley scattering. Consider the scattering amplitude
\begin{eqnarray}
& \langle K'{\bf p}' | \hat{V} | K{\bf p}\rangle =
{\textstyle \frac{\hbar\lambda va}{2iS} \sin\frac{\varphi_{\bf p}+\varphi_{\bf p'}}{2}}
\sum_l e^{i\theta_l}, \label{eq:ampl_sum} \\
& \theta_l = \frac{2\pi m_l}{3} + ({\bf p-p'}) {\bf r}_l. \nonumber
\end{eqnarray}
At temperatures far from $T_c$, $T\gg T_c$, adatom positions on the superlattice are random so that
$m_l$ take values $-1, 0$ and $1$ with equal probabilities.
As a result, the absolute value of scattering amplitude can be estimated as
$|\langle K'{\bf p}' | \hat{V} | K{\bf p}\rangle| \sim \sqrt{n_i\lambda^2_F}$.
Upon approaching $T_c$ from above, clusters of ordered adatoms with a characteristic size $\xi\gg n^{-1/2}_i$
start appearing. In the sum (\ref{eq:ampl_sum}) such a cluster generates constructive interference
between terms with the same value of $m_l$ provided that $\xi\lesssim \lambda_F$.
This increases the scattering amplitude,
$|\langle K'{\bf p}' | \hat{V} | K{\bf p}\rangle| \sim n_i\xi\lambda_F$.
A further increase of the correlation length, $\xi>\lambda_F$, has an opposite effect on
scattering: electrons get scattered only by the gradients in the smoothly fluctuating field
$\overline{\bf u}$.

The intervalley momentum relaxation rate (both at $T<T_c$ and $T>T_c$) \cite{BornApplicability}
$\tau^{-1}_i$ can be expressed in terms of the Fourier transform of the correlation function,
$\tilde{g}(k)=\int d{\bf r} \, e^{i{\bf kr}} g(r)$,
\begin{eqnarray}&
\displaystyle \frac{1}{\tau_i} = \frac{\hbar v^2 p^2_F n_i\lambda^2 a^2}{2 \varepsilon_F}
\int\limits^{2\pi}_0 \frac{d\varphi}{2\pi} (1-\cos\varphi) r_i(\varphi), \label{eq:tau_i} \\
& r_i(\varphi) = \displaystyle  \left[1 + n_i \tilde{g}\left(2p_F \sin\frac{\varphi}{2}\right)\right]
\left[ 2\sin^2\frac{\varphi}{2} + \frac{\Delta^2(T)}{(\hbar vp_F)^2} \right]. \nonumber
\end{eqnarray}

At $T>T_c$ ($\Delta=0$) temperature dependence $\rho(T)$ comes from the correlation function $\tilde{g}(k)$ in
Eq.~(\ref{eq:tau_i}). Far from the phase transition, $|T-T_c|\sim T_c$, where $\xi<\lambda_F$
(region III in Fig.~\ref{fig:resist}), electrons are effectively scattered by small clusters of ordered adatoms.
In this region we approximate $\tilde{g}(k)\approx\tilde{g}(0) \propto (\sqrt{n_i}\xi)^{2-\eta}$
and find that
\begin{equation}
\delta\rho(T) \approx C \frac{\Delta^2(0) }{e^2 v^2 \hbar n^{\eta/2}_i} \xi^{2-\eta}
\propto \frac{n^{2-\eta/2}_i}{(T-T_c)^{(2-\eta)\nu}},
\end{equation}
where $C=(3\pi^2/2)\int_0^{+\infty} dy\, y^{1-\eta} \kappa(y)$ is a dimensionless constant.

In the vicinity of the critical point, such that $\lambda_F<\xi$ (region II in Fig.~\ref{fig:resist}),
electrons experience multiple scatterings within one cluster with a small wave vector transfer, $\sim\xi^{-1}$.
This makes $\rho(T)$ sensitive to the critical behaviour of the
correlation function at $r<\xi$. This region is easier to analyze
by performing the angular integration in Eq.~(\ref{eq:tau_i}) and expressing $\tau^{-1}_i$
in terms of the function $\kappa(y)$ defined in Eq.~(\ref{eq:corr_fnc}):
\begin{equation}
\displaystyle \frac{1}{\tau_i} = \frac{\Delta^2(0)}{\hbar^2 v} \left[\frac{3p_F}{4n_i} +
\frac{2\xi}{(\sqrt{n_i}\xi)^\eta} \int\limits^{+\infty}_{0} dy\,
\displaystyle \frac{\kappa(y)}{y^\eta} f(p_F\xi y) \right].
\label{eq:rho_i}
\end{equation}
Here, $f$ can be expressed in terms of Bessel functions as
\begin{equation}
f(x) = \frac{\pi}{x} [x^2 J^2_0(x) - x^2 J^2_1(x) + J^2_1(x) - x J_0(x) J_1(x)].
\nonumber
\end{equation}
To evaluate the integral in Eq.~(\ref{eq:rho_i}) we divide the integration interval $[0,\infty]$ into two parts,
$[0, y_0]$ and $[y_0,\infty]$, where $1\gg y_0\gg 1/p_F\xi \to 0$.
For the interval $[y_0,\infty]$, we use the fact that $f(p_F\xi y)$
is a fast oscillating function, $f(x\gg 1)\approx 2\sin(2x)+ 3\cos(2x)/(2x)$,
and that
\begin{equation}
\int\limits^{+\infty}_{y_0} F(y) \sin(Ay) dy = \frac{F(y_0) \cos(A y_0)}{A} + O\left(\frac{1}{A^2}\right),
\nonumber
\end{equation}
for $A\gg 1$ and $F(\infty) = 0$. For the interval $[0,y_0]$ in the leading order in
$1/p_F\xi$, the result is determined by the values of $\kappa_1(0)$ and $\kappa_2(0)$ in Eq.~(\ref{eq:correl_above}).
For this, we expand $\kappa_1$ and $\kappa_2$ (which vary at the scale of $y\sim 1$) into Taylor series, evaluate 
the corresponding integrals in the leading orders in $y_0\ll 1$, and combine with the contribution
from the interval $[y_0, \infty]$. As a result, the term with $\kappa_1$ in Eq.~(\ref{eq:rho_i})
produces a finite contribution when $p_F\xi\to \infty$, and we find that
\begin{equation}
\rho(T_c) - \rho(\infty) = \frac{ c_1 \pi^{\frac{3}{2}} \Delta^2(0)}{e^2v^2 \hbar n_i}
B(\eta) \left(\frac{n_i}{\pi n_e}\right)^{1-\eta/2},
\end{equation}
where $B(x) = x\Gamma\left(\frac{3+x}{2}\right) \Gamma\left(1-\frac{x}{2}\right)/
\Gamma\left(1+\frac{x}{2}\right) \Gamma\left(2+\frac{x}{2}\right)$.
The next term in the expansion $\kappa_1(y) = \kappa_1(0) + y \kappa'_1(0) +\ldots$ generates a contribution
$O(1/p_F\xi)$, which, for $T\to T_c$, is less relevant than a more singular $(T-T_c)$-dependent \cite{CriticalExp}
contribution from the $\kappa_2$ term in Eq.~(\ref{eq:rho_i}). Following to same steps,
we find that the latter term gives rise to the cusp in the $\rho(T)$ dependence
in the region II near $T_c$ in Fig.~\ref{fig:resist},
\begin{equation}
\frac{\rho(T)-\rho(T_c)}{\rho(T_c)-\rho(\infty)} = 
- \frac{c_2 B(\eta-2\gamma)}{c_1 B(\eta)(\pi n_e \xi^2)^{\gamma}}
\propto (T-T_c)^{1-\alpha},
\label{eq:rho_nonanalytic}
\end{equation}
where $\gamma=(1-\alpha)/2\nu$.

The behaviour of $\delta\rho(T)$ at $T<T_c$ (region I in Fig.~\ref{fig:resist}) is determined by two contributions.
One part, $\delta\rho_1/\rho \propto (T_c-T)^{1-\alpha}$,
is related to the specific heat anomaly correction to the correlation function
and can be obtained the same way as Eq.~(\ref{eq:rho_nonanalytic}).
The other contribution, $\delta\rho_2/\rho \propto (T_c-T)^{2\beta}$,
is due to the formation of a non-zero order parameter in the Kekul\'{e}-ordered phase.
The second correction dominates when $2\beta<1-\alpha$, which is the case
for the expected values of critical exponents \cite{CriticalExp}.
As a result, we attribute the rise of resistivity at $T<T_c$ near the cusp at $T=T_c$
to the formation of a spectral gap in graphene due to the Kekul\'{e} mosaic ordering.
The resulting behaviour of resistivity as a function of temperature for all three regimes
is plotted in Fig.~\ref{fig:resist} for $c_1=0.5$, $c_2=0.15$,
$\varepsilon_F=0.4v\sqrt{n_i}$, $n_i\lambda^2a^2=0.005$ \cite{CriticalExp}.

In conclusion, we investigated electron transport in graphene covered
by a dilute ensemble of adatoms residing over the centers of hexagons.
We calculated the temperature dependence of the resistivity $\rho(T)$,
which appears to be non-monotonic and has a non-analytic cusp at $T=T_c$.
Since the form of the cusp depends on the critical indices $\alpha$ and $\beta$
of the phase transition, experimental observation of such anomaly may facilitate their measurements.
The form of $\rho(T)$ shown in Fig.~\ref{fig:resist} appears to be
generic for partially ordered dilute ensembles of adatoms with alternative positioning on the honeycomb lattice:
(i) over the sites \cite{Sublattice,Levitov} and (ii) over carbon-carbon bonds \cite{AllTypes}.
Since those also fall into the class of Potts models [(i) - 2 value Potts model and (ii) - 3 value Potts model],
the anomalies in $\rho(T)$ can be described using Eqs.~(\ref{eq:rho_delta},\ref{eq:rho_nonanalytic})
with appropriate critical indices, $\alpha$ and $\beta$.

This work was supported by EPSRC under Grant Nos. EP/G041954 and EU-FP7 ICT STREP Concept Graphene,
US DOE contract No.DE-AC02-06CH11357 and The Royal Society.


\begin{thebibliography}{99}
\bibitem{CriticalExp} Critical exponents, satisfying the relations $2\nu=2-\alpha$, $2\beta=\nu\eta$,
are unknown for disordered Potts model. So that we use the exponents of
three-state Potts model on a square lattice as a reference: $\alpha=1/3$, $\beta=1/9$, $\nu=5/6$, $\eta=4/15$ \cite{Baxter}.
The small universal corrections to these values due to weak disorder were derived by Dotsenko \cite{Dotsenko}.
To mention, 3-state Potts model with $1/r^3$ interaction in 2D without disorder
features a first-order phase transition, however, disorder makes it the second order
transition \cite{AizenmanWehr}.
\bibitem{BornApplicability} The Born approximation is applicable for this problem only if electron energy is larger than
the average spectrum gap at the length scale of $\lambda_F$, $\Delta(\lambda_F) < \hbar vp_F$, where
\begin{equation}
\begin{array}{c}
\Delta(L) = \hbar n_i \lambda v a \sqrt{\overline{{\bf u}({\bf r}) {\bf u}({\bf r'})}_L} \sim \\
\displaystyle\frac{\hbar \sqrt{n_i} \lambda va}{L} (\sqrt{n_i} \text{min}(\xi, L))^{1-\eta}, \\
\end{array}
\nonumber
\end{equation}
so that
$\text{min}(\xi,\lambda_F) < \frac{1}{\sqrt{n_i}} \left(\frac{4\pi^2}{n_i \lambda^2 a^2}\right)^{1/(2-\eta)}$,
which coincides with the inequality $v p_F \tau_i \gg 1$, that is, if graphene resistivity is smaller than $h/e^2$.
\end{thebibliography}
\end{document}